\documentclass[journal,comsoc,onecolumn,12pt]{IEEEtran}
\usepackage[normalem]{ulem}

\usepackage{array}
\usepackage{algorithmic}
\usepackage{textcomp}
\usepackage{nicefrac}
\usepackage{graphicx}
\usepackage{amsmath,amsfonts,amssymb,latexsym}
\usepackage{diagram}
\usepackage{tikz}
\usetikzlibrary{shapes,arrows,positioning,fit}
\usepackage{macro}
\usepackage{arydshln}

\newcommand{\upsample}[1]{\mbox{$\uparrow #1$}}
\newcommand{\upsampler}{\mbox{$\upsample{M}$}}
\newcommand{\hold}[1]{\mbox{${\cal H}_{#1}$}}
\newcommand{\holdM}{\mbox{$\hold{h/M}$}}
\newenvironment{caseseq}{%
	\left\{\begin{array}{@{}l@{}>{{}}l@{}}
}{%
	\end{array}\right.
}

\graphicspath{{./Figs/}}

\tikzset{
	arrow/.style = {
		shorten >= 2pt,
		shorten <= 2pt,
	},
}
\usepackage{etoolbox}
\AtEndEnvironment{Example}{\hfill$\diamondsuit$}

\begin{document}
\title{Hypertracking and Hyperrejection: Control of Signals beyond the Nyquist Frequency}
\author{Kaoru Yamamoto, Yutaka Yamamoto, and Masaaki Nagahara
\thanks{K. Yamamoto has been supported in part by the Japan Society
for the Promotion of Science under Grants-in-Aid for Scientific
Research No.\ 20K14766. K. Yamamoto and Y. Yamamoto have
been supported in part by the Japan Society for the Promotion of
Science under Grants-in-Aid for Scientific Research No.\ 19H02161.
Y. Yamamoto wishes to thank DIGITEO and Laboratoire des Signaux et
Systemes (L2S, UMR CNRS), CNRS-CentraleSupelec-University Paris-Sud
and Inria Saclay for their financial support while part of this research
was conducted. M. Nagahara has been supported in part by the Japan Society
for the Promotion of Science under Grants-in-Aid for Scientific Research
No.\ 20H02172.}
\thanks{K. Yamamoto is with the Faculty of Information Science and Electrical Engineering, Kyushu University, Fukuoka 819-0395, Japan (e-mail: yamamoto@ees.kyushu-u.ac.jp). }
\thanks{Y. Yamamoto is Professor Emeritus, Graduate School of Informatics,
Kyoto University, Kyoto 606-8510, Japan (e-mail: yy@i.kyoto-u.ac.jp).}
\thanks{M. Nagahara is with the Institute of Environmental Science and Technology,
the University of Kitakyushu, Fukuoka 808-0135, Japan (e-mail: nagahara@ieee.org).}}

\maketitle

\begin{abstract}
This paper studies the problem of signal tracking and disturbance
rejection for sampled-data control systems, where the pertinent
signals can reside beyond the so-called Nyquist frequency.
In light of the sampling theorem, it is generally understood
that manipulating signals beyond the Nyquist frequency
is either impossible or at least very difficult.  On the other hand,
such control objectives often arise in practice, and
control of such signals is much desired.
This paper examines the basic underlying assumptions in the
sampling theorem and pertinent sampled-data control schemes,
and shows that the limitation above can be removed
by assuming a suitable analog signal generator model.
Detailed analysis of multirate closed-loop systems,
zeros and poles are given, which gives rise to tracking
or rejection conditions.  Robustness of the
new scheme is fully characterized; it is shown that there is a
close relationship between tracking/rejection frequencies
and the delay length introduced for allowing better
performance.
Examples are discussed to illustrate the effectiveness of the
proposed method here.
\end{abstract}

\section{Introduction}

It is well recognized that sampled-data control systems 
are inherently limited in resolution in time, due to 
sampling.  This is clearly seen from the classical 
sampling theorem, e.g., \cite{Shannon1}; there 
Shannon showed that 
the original analog signal can be 
perfectly reconstructed if the signal is perfectly 
band limited below the so-called Nyquist frequency, 
i.e., half the sampling frequency.  

In spite of such well-established developments, there 
are many practical needs to process signals that 
go {\em beyond the Nyquist frequency}.  
Superresolution in image processing is one such example.
In control as well, we are often confronted with 
such requirements.  
For example, 
\begin{itemize}
\item tracking a high-frequency sinusoid in 
regulating AC power current to a prespecified 
frequency (particularly in micro-grid systems), 
\item rejection of high-frequency disturbance
generated by disk rotation in hard-disk drives
\cite{Atsumi2010,Zhengetal2016}, or 
\item laser sintering manufacturing systems \cite{Frazier2014,Xiaoetal2018}, 
vision-guided high-speed controls \cite{Tanietal2014}, etc. 
\end{itemize}

Due to the limited resolution in time, such 
objectives have been regarded as either impossible or at 
least ill posed \cite{YamamotoEnc2018}. 
However, if we examine the sampling theorem, it is clear that
the band-limitation below the Nyquist frequency is only 
a {\em sufficient\/} condition for perfect signal reconstruction.  
A different type of band-limiting 
hypothesis can lead to a different signal reconstruction 
result; see, e.g., \cite{Sagarfest2017}.  
Indeed, by using modern sampled-data control theory, 
we have developed a new paradigm 
for digital signal processing, including superresolution
\cite{YYNagPPKSP12}. 

The present work is motivated by the above observation, 
and intends to give a solution to the control problems 
as listed above.  
More specifically, we study high-frequency tracking and disturbance rejection
of signals beyond the Nyquist frequency.  
The basic philosophy remains the same as that of \cite{YYNagPPKSP12}, 
with the differences that 
the tracking or rejection signals must be precise, 
and we must also form a closed-loop system.  Robustness 
becomes a crucial issue here.  

In view of the new feature of tracking or rejecting signals
beyond the Nyquist frequency, we call the present control 
scheme {\em hypertracking\/} or {\em hyperrejection\/} to 
highlight the difference with conventional tracking/rejection 
problems in sampled-data control.  

The basic concept of the present study for hypertracking 
was first presented in \cite{YYNCDC16}.  
This paper is a continuation 
with a complete characterization of 
robustness, which gives a new insight 
(see Section V) to us.  
A related study concerning multiple 
signals was also given in \cite{KYYYNagCCTA17} in a different 
setting. 
We also note that the recent article \cite{Xiaoetal2018} 
has proposed a multirate control scheme to reject 
the disturbance beyond the Nyquist frequency in a 
mechatronic system. However, the optimized intersample behavior
and the robustness are not addressed there. 


\subsection*{Notation}
In denoting function values, we will adopt the following convention:
for a function $f$ with a continuous-time variable $t$, we 
write $f(t),$ while we write $g[k]$ with square brackets when 
$k$ takes on integer values.

\section{Problem Formulation}
  \label{Sec:formulation}

\begin{figure}[htb]
\centering
\tikzstyle{block} = [draw,  rectangle, minimum height=0.5cm, minimum width=0.4cm]
\tikzstyle{sum} = [draw, circle, node distance=0.7cm]
\tikzstyle{input} = [coordinate]
\tikzstyle{output} = [coordinate]
\tikzstyle{disturbance} = [coordinate]
\tikzstyle{pinstyle} = [pin edge={to-,thin,black}]
\begin{tikzpicture}[auto, node distance=1.4cm,>=latex']
\centering
    \node [input, name=input]{};
    \node [sum, right of=input] (sum1) {};
    \node [block, right = 0.9cm of sum1] (upsampler) {$\upsampler$};
    \node [block, right of=upsampler] (controller) {$K(z)$};
    \node [block, right of=controller] (hold){$\mathcal{H}_{h/M}$};
    \node [sum, right = .5cm of hold] (sum2) {};
    \node [block, right = .5cm of sum2] (plant)	{$P(s)$};
    \node [disturbance, above = .8cm of sum2] (disturbance){};
    \node [output, right of=plant] (output){};
    \node [coordinate, below of=controller] (belowK){};
    \draw [draw,->] (input) --node{$r$}(sum1);
    \draw [-] (sum1) --node{$e$} (1.2,0);
    \draw [-] (1.2,0) -- node[xshift=0.3cm,yshift=0.1cm]{$h$}(1.6,0.2);
    \draw [->] (1.6,0)-- (upsampler)--(controller);
    \draw [->] (controller) -- (hold);
    \draw [->] (sum2) -- (plant);
    \draw [->] (hold) -- node[pos=0.99]{$+$}(sum2);
    \draw [->] (disturbance) -- node{$d$}(sum2);
    \draw [->] (plant) -- node[name=y]{$y$}(output);
    \draw [->] (y) |-(belowK) -|  node[pos=0.95,xshift=0.5cm]{$-$} (sum1); 
 \end{tikzpicture}
 \caption{Sampled-data feedback system with input disturbance}
\label{Fig:1}
\end{figure}
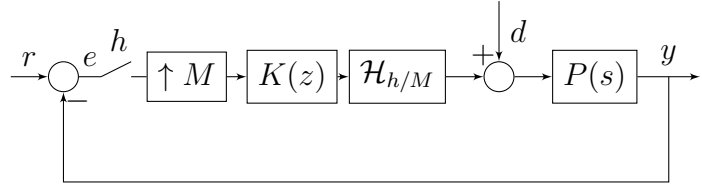

Consider the sampled-data system depicted in
Fig.~\ref{Fig:1}.

\noindent
$P(s)$ is a linear, time-invariant,
strictly proper,
continuous-time plant,
and $K(z)$ is a linear, time-invariant, discrete-time controller.
The error $e$ is sampled with sampling period $h$, and
after sampled, it is upsampled by factor $M$ to allow for a
faster control processing.  The role of the action
({\em upsampler\/}) of $\upsampler$ is to increase the
sampling rate by the factor $M$ by placing $M-1$ equally
spaced zeros between each pair of samples as follows (\cite{Fliege,Vaidyanathan}):
\begin{equation*} 
  (\upsampler)(e)[k] =
    \begin{cases}
       e[k/M] & \mbox{ if $k = mM$ for some integer $m$} \\
       0      & \mbox{ otherwise}.
    \end{cases}
\end{equation*}

Note that in the present case, the sampled sequence $\{e[k] \}_{k=1}^{\infty}$
synchronizes with continuous-time signals
every $h$
seconds, hence $e[k]$ enters into the system as $e(kh)$ for every $k$.
With this convention, the action of the above upsampler takes the
following form:
\begin{equation} \label{eqn:upsamplerdef}
  (\upsampler)(e)(kh+\ell)=
    \begin{cases}
       e(kh) & \text{if }\ell = 0 \\
       0     & \text{if }\ell = \nicefrac hM, \ldots, \nicefrac{(M - 1)h}{M}.
    \end{cases}
\end{equation}
$\holdM$ is the zero-order hold that holds the output
as constant for the period of $h/M$.

We now consider the following problem:

\noindent
\textbf{Problem 1}: \emph{In the block diagram Fig.~\ref{Fig:1},
consider the reference input $\sin \omega_{r} t$ and
the disturbance input $\sin \omega_{d} t$
where $\omega_{r}$ and $\omega_{d}$ are either
below or above the Nyquist frequency
$\pi/h$.  Find a discrete-time controller $K(z)$
such that the output $y$ that tracks the reference
$r(t) = \sin \omega_{r} t$ or its delayed signal $r(t-L)$
for some $L > 0$, and also rejects the disturbance $d(t)$.
Here the tracking may be only approximate due to the
sample-hold nature in Fig.~\ref{Fig:1}.  However,
the error due to this approximation becomes small
for a large $M$ or small $h$ (\cite{YYAutoma99}).}  

When $\omega_{r}$ or
$\omega_{d}$ is above the Nyquist frequency, this problem
does not fall into the conventional sampled-data control
paradigm.  Such signals appear in measurement as
aliased components below the Nyquist frequency, and
are mixed with other system signals already
existent in the base-band (i.e., lower than the Nyquist frequency)
range.  The prime objective of the present paper is to provide
a scheme that enables us to achieve the above goal.

\section{Hypertracking and Tracking Conditions}
  \label{Sec:Models}

We now proceed to give a solution
to the hypertracking problem.  We first give a state-space
description of Fig.~\ref{Fig:1}, assuming $d\equiv 0$, characterize
its zeros for tracking, and then proceed to a design method
and examples in the subsequent sections.

\subsection{State-space description of the lifted multirate system}
\label{Sec:LiftedSS}
We first describe the system in Fig.~\ref{Fig:1} as a time-invariant
discrete-time system with a single sampling period $h$.

Let $P(s)$ and $K(z)$ be described by the following
state-space equations:
\begin{align}
&P(s):
   \begin{caseseq}
      \frac{d}{dt} x_{c}(t) & =  A_{c}x_{c}(t) + B_{c}u(t) \\
      y(t) & =  C_{c}x_{c}(t)
   \end{caseseq}\nonumber\\  
&K(z):
   \begin{caseseq}
      x_{d}[k+1] & =  A_{d}x_{d}[k] + B_{d}w_{d}[k] \\
      y_d[k] & =  C_{d}x_{d}[k] + D_{d}w_{d}[k].
   \end{caseseq} \label{eqn:ctrlss}
\end{align}
Here $x_{c}$, $y$, $u$ denote, respectively, the state, output
and input of the plant $P(s)$, and $x_{d}$, $y_{d}$, $w_{d}$
the state, output and input of the controller $K(z)$.
Note that the discrete-time controller \eqref{eqn:ctrlss}
operates in conformity with the sampling period $h/M$.
That is, $x_d[k], y_d[k]$ and
$w_d[k]$ occur at time $t=kh/M,$ $x_d[k+1], y_d[k+1], w_d[k+1]$
at $t=(k+1)h/M$, and so on.

Introduce the continuous-time
lifting \cite{ChenFrancisBook95,BPFT91,BamiehPearson92,YYTAC94}:
\begin{equation*} 
\begin{split}
  \mathcal{L}:\, & \Ltwoloc \rightarrow \ell^{2}(L^{2}[0, h)):
               x(\cdot) \mapsto \{x[k](\cdot) \}_{k=0}^{\infty}, \\
            &    x[k](\theta) := x(kh+\theta), \ \theta \in [0,h).
 \end{split}
\end{equation*}

We then have the following:
\begin{Prop} \label{Prop:disclifting} 
\em
When lifted with period $h$, the closed-loop system Fig.~\ref{Fig:1},
without disturbance $d$, is described by
\begin{equation}
\begin{bmatrix}
  \bar x_d[k+1] \\ x_{c}[k+1]
\end{bmatrix}
   = \begin{bmatrix} \bar{A_d} & -\bar{B_d}C_c \\
	  B(h)\bar{C_d} & e^{A_ch}-B(h)\bar{D_d}C_c
      \end{bmatrix}
      \begin{bmatrix}
          \bar x_d[k] \\ x_{c}[k]
      \end{bmatrix}
     + \begin{bmatrix}
           \bar{B_d}\delta_0 \\ B(h)\bar{D_d}\delta_0
       \end{bmatrix}r[k](\theta)
            \label{eqn:clloopeqn}
\end{equation}
and
\begin{align}
e[k](\theta)
        &=r[k](\theta)-y[k](\theta) \nonumber \\
	&= \begin{bmatrix}
            -C_cB(\theta)\bar{C_d} & -C_ce^{A_c\theta}+C_cB(\theta)\bar{D_d}C_c
           \end{bmatrix}
           \begin{bmatrix}
              \bar x_d[k] \\ x_{c}[k]
           \end{bmatrix}
+ (I-C_cB(\theta)\bar{D_d}\delta_0)r[k](\theta),
             \label{eqn:ek}
\end{align}
where
\begin{align}
    &\bar{x}_d[k]:=x_d[kM], \quad
  	\bar{y}_d[k]:=\begin{bmatrix}
                   y_d[kM] \\ y_d[kM+1] \\ \vdots \\y_d[(k+1)M-1]
               \end{bmatrix}, \nonumber\\[0.8em]
         & \bar{A_{d}} :=A_d^M,\quad \bar{B_{d}} := A_d^{M-1}B_d, \quad
		\bar{C_{d}}  :=\begin{bmatrix}
		   C_d \\ C_dA_d \\ \vdots \\ C_dA_d^{M-1}
	         \end{bmatrix},
	         \quad
		\bar{D_{d}} :=
		\begin{bmatrix}
		   D_d  \\C_dB_d \\ \vdots  \\C_dA_d^{M-2} B_d
		\end{bmatrix}, \nonumber\\[0.8em]
&H(\theta) := [\chi_{[0, \frac{h}{M})}(\theta),
              \chi_{[\frac{h}{M}, \frac{2h}{M})}(\theta),
              \dots,
              \chi_{[\frac{(M-1)h}{M}, h)}(\theta)],\label{eqn:genhold}
\end{align}
with $\chi_{[\frac{ih}{M}, \frac{(i+1)h}{M})}(\theta)$, $i=0, \dots, M-1$
being the characteristic function of the interval $[ih/M, (i+1)h/M)$,
\begin{equation}
  B(\theta) :=
        \int_0^{\theta} \! e^{A_c(\theta-\tau)}B_cH(\tau) ~ \mathrm{d}\tau,
\end{equation}
and $\delta_{0}$ denotes Dirac's delta, acting on $r[k](\theta)$ as
$\delta_{0} r[k](\theta) := r[k](0)$.
\end{Prop}
\rm

\Proof
Direct calculation.  See \cite{YYNCDC16} for details;
see also \cite{QiuChen1999,Voulgarisetal1994} for
pertinent calculation.
\EndProof

\subsection{Zeros and tracking}
\label{Sec:Zeros}
As discussed in \cite{YYTAC94}, the tracking performance of 
the closed-loop system (\ref{eqn:clloopeqn}), (\ref{eqn:ek}) 
is determined by 
\begin{enumerate}
\item the transmission zeros, and 
\item the corresponding zero directions, each of which is the initial 
      intersample function of the tracking signal.
\end{enumerate}

We make the following assumption: 

\noindent
\textbf{Assumption A}: There is no pole-zero cancellation 
between the lifted discrete-time controller and 
the continuous-time plant.  

The following theorem has been obtained in 
\cite{YYNCDC16}: 
\begin{Thm} \label{Thm:transmissionzero} 
\em
Under Assumption A, and the assumption of the closed-loop
stability, the unstable poles of $K$ and lifted $P$ 
induce a transmission zero of the closed-loop transfer operator
$G_{er}(z)$ and vice versa. 
\end{Thm}
\rm

As a corollary, consider the case 
$\lambda = e^{j\omega h}$ being an eigenvalue of $\bar{A_{d}}$.  
Let $\bar{x}_{d}$ be the corresponding eigenvector, and then 
$\bar{C_{d}}\bar{x}_{d} \neq 0$ (otherwise, the controller is not 
observable).  It follows that by taking $H(\theta)$ to be 
$e^{j\omega \theta}$ ($0\leq \theta \leq h$), the output 
of the discrete-time controller becomes 
$e^{jk\omega h}e^{j\omega \theta}\bar{y}_{d}
= e^{j\omega t}\bar{y}_{d}$, where $\bar{y}_{d} = \bar{C_{d}}\bar{x}_{d}$.  
That is, the discrete-time controller can work as an internal
model for $1/(s - j\omega)$.  Taking a combination with 
the complex conjugate, this can work as an internal model for 
$\sin \omega t$, with this suitable choice of $H(\theta)$.  
When $H(\theta)$ given by \eqref{eqn:genhold}
is the zero-order hold, it cannot exactly 
produce this sinusoidal hold function, but it can still approximate 
such a hold function.  

\begin{Remark} \label{Rem:intmodel} \rm
In fact, if $e^{j\omega h}$ is an eigenvalue of $\bar{A_{d}} = A_{d}^{M}$, 
$e^{j\omega h/M}$ is an eigenvalue of $A_{d}$.  Then, by 
taking ${\cal H}_{h/M}$ to be $e^{j\omega \theta}$ for 
$0 \leq \theta \leq h/M$, it is seen that the output of the controller 
produces $e^{j\omega t}$, because at each step the output is kept 
multiplied by $e^{j\omega h/M}$.  
Since the difference between the zero-order hold and the sinusoidal 
hold $e^{j\omega \theta}$ is small for $0 \leq \theta \leq h/M$, 
for a sufficiently large $M$,
the controller output can produce an approximation of $e^{j\omega t}$.
This indeed occurs in the subsequent 
Fig.~\ref{Fig:controloutput} in Example \ref{Ex:0.75Hz}.  
\end{Remark}

\section{Design Method}
   \label{Sec:Design}
We now proceed to give a solution to Problem 1.
However, Fig.~\ref{Fig:1} as it is cannot be
used as a design block diagram for sampled-data $\Hinf$
control since sampling is not a bounded operator on $L^{2}$.
We thus place a strictly proper anti-aliasing filter
$F(s)$ in front of the adding point of the error.  In other words,
the reference signal is pre-filtered by $F(s)$.  This is
also advantageous in that we can control frequency weighting in
the input reference signals, which plays a crucial role
in our hypertracking problem.
{\em Unlike the usual case, we
place more emphasis on the frequency that we wish to track,
possibly beyond the Nyquist frequency}.
While we confine our discussions to tracking problems,
it is straightforward to see that disturbance rejection can be treated
in exactly the same way.

We also allow some delays in
tracking.  Instead of taking the error
$e(t) = r(t) - y(t)$, we try to minimize the delayed error
$\tilde{e}(t) := r(t-L) - y(t)$ for some $L > 0$ as
stated in Problem 1.  This will give us extra freedom
in designing our controller.
On the other hand, we will also
see that this delay places certain limitations in robustness; see
Section \ref{Sec:robustness} below.

\begin{figure} [h]
\centering
\begin{tikzpicture}
\node[draw,minimum size=2cm] (GP) {$\begin{bmatrix}  e^{-Ls}F &  -P\\
										F &  -P \end{bmatrix}$};
\draw[arrow, <-] ([xshift=-1cm,yshift=-10pt]GP.west) -- node[yshift=0.15cm]{$e$}+(1cm,0pt);
\draw[arrow, <-] ([xshift=-1cm,yshift=10pt]GP.west) --  node[yshift=0.15cm]{$\tilde{e}$} +(1cm,0pt);
\draw[arrow, <-] ([yshift=-10pt]GP.east) -- node[yshift=0.15cm]{$u$} +(1cm,0pt);
\draw[arrow, <-] ([yshift=10pt]GP.east) -- node[yshift=0.15cm]{$r$} +(1cm,0pt);
\end{tikzpicture}
 \caption{Generalized plant}
 \label{Fig:genplant}
\end{figure}
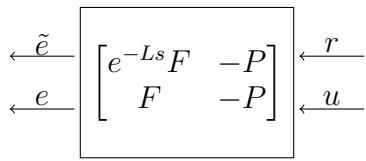

Incorporating these changes into Fig.~\ref{Fig:1}, we
obtain the generalized plant in Fig.~\ref{Fig:genplant}
for design.  Here $L$ is a design parameter; we usually take
$L$ to be an integer multiple of $h$, with some small number
such as $4$--$10$.  Problem 1 is now restated
as the following sampled-data $\Hinf$-control design problem:

\noindent
\textbf{Problem 1a:} \emph{Given $F(s), P(s), L$ and an attenuation
level $\gamma > 0,$ find a stabilizing digital controller $K(z)$ such that
\[
 \|G_{\tilde{e}r}(s)\|_\infty < \gamma
\]
where $G_{\tilde{e}r}(s)$ denotes the transfer operator from
$r$ to $\tilde{e}.$}

The solution to this sampled-data $\Hinf$-control problem can be
obtained via standard solutions; see, e.g., \cite{ChenFrancisBook95}, \cite{Kahaneetal1999},
\cite{YamamotoEnc2018} and references therein.
The only nonstandard element is the
delay $e^{-Ls}$, which is an infinite-dimensional operator.
It is quite effective to rely on the
fast-sample/fast-hold approximation introduced by
\cite{AndersonKeller1998}.  See also \cite{YYAutoma99} for
pertinent discussions.

We start with the simplest case of hypertracking:
\begin{Example} \label{Ex:0.75Hz}\rm
Consider the plant
\begin{equation}\label{eqn:plant}
  P(s) := \frac{1}{s^{2} + 2s + 1}
\end{equation}
with (normalized) sampling period $h = 1$ in Fig.~\ref{Fig:1}.
From here on, we always normalize the sampling period $h$ to 1.
The Nyquist frequency is then $\pi$ [rad/sec] which is just
equal to $0.5$ [Hz].
Suppose that we are given
the tracking signal $r = \sin (3\pi/2)t$,
i.e., the sinusoid at $0.75$ [Hz].
This is clearly above the Nyquist frequency, and a normal
signal-processing intuition or a digital control thinking
may claim that it is impossible to track.

The basic idea is that we place more weight on this high
frequency signal rather than the low frequency range below
the Nyquist frequency.  In this example, we take the weighting
function
\begin{equation*} \label{eqn:weight1}
  F(s) := \frac{s}{s^{2} + 0.1s + (3\pi/2)^{2}},
\end{equation*}
which has a sharp peak at $3\pi/2$ [rad/sec] and also deemphasizes
low-frequency as can be seen in Fig.~\ref{Fig:F075}.

\begin{figure}[tb]
\centering
\includegraphics{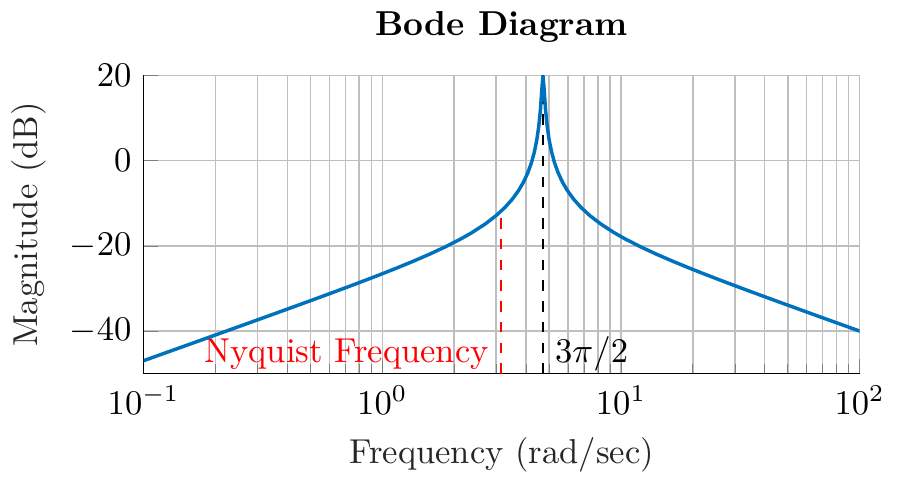}
\caption{Weighting function $F(s)$ in Example~\ref{Ex:0.75Hz}}
\label{Fig:F075}
\end{figure}

The response against the sinusoid $r(t) = \sin (3\pi/2)t$ is shown in
Fig.~\ref{Fig:output0.75} along with the delayed error, represented by
the dashed line. Here we chose the upsampling factor $M=8$ and the delay $L=4h=4$.
\begin{figure}[tbp]
\centering
    \includegraphics{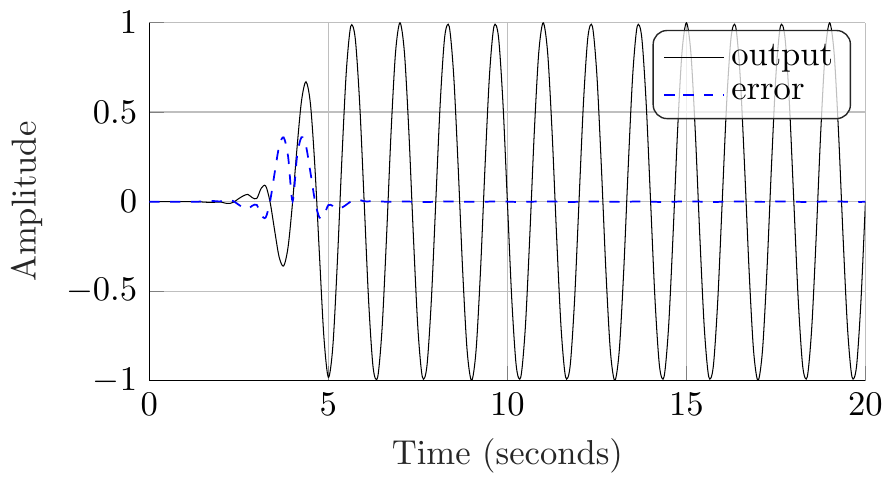}
\caption{System output tracking $\sin (3\pi/2)t$ along with the delayed error}
\label{Fig:output0.75}
\end{figure}
This figure clearly shows that the output tracks the
reference input $\sin (3\pi/2)t$, which has the natural
frequency greater than the Nyquist frequency $\pi$, and the
output matches the given frequency $3\pi/2$.  Note also
that the output shows the delay of $4$ steps
specified by the design specification.
\end{Example}

\begin{figure}[tbp]
\centering
    \includegraphics{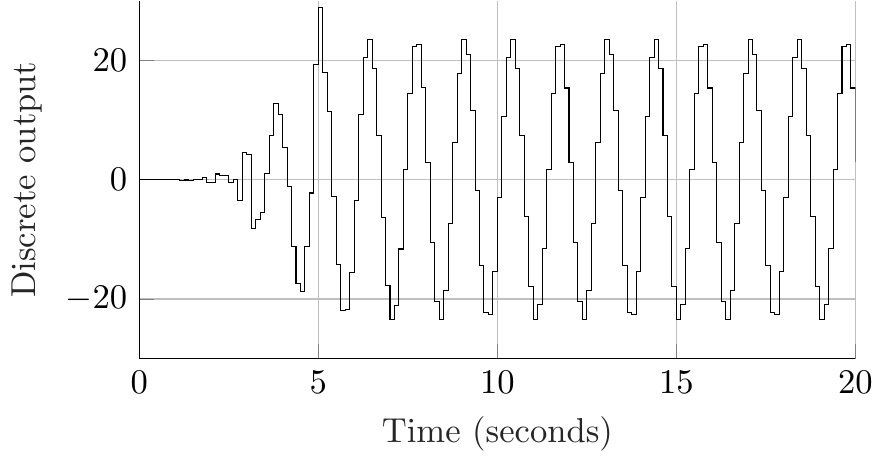}
\caption{Discrete-time controller output}
\label{Fig:controloutput}
\end{figure}

\begin{figure}[t]
\centering
    \includegraphics{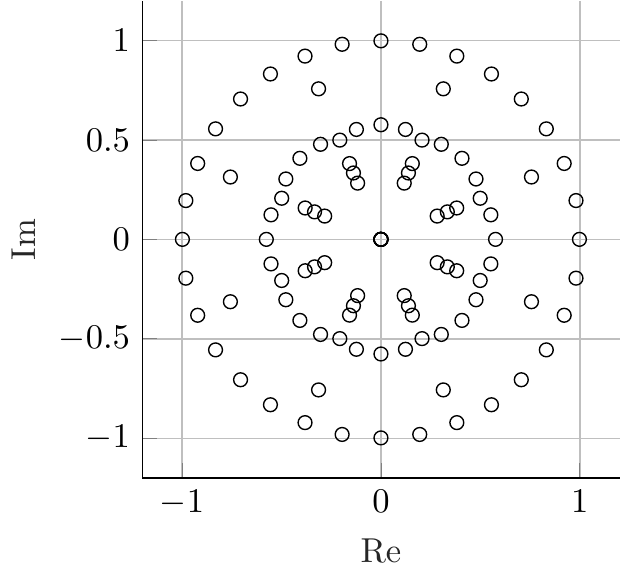}
\caption{Poles of the controller}
\label{Fig:poles}
\end{figure}
Fig.~\ref{Fig:poles} shows
the eigenvalues of the upsampled controller, i.e., those of $A_{d}$.
There are poles at $\pm j$ corresponding to
$e^{\pm 3\pi j/2}$---necessary to produce $\sin (3\pi/2)t$,
along with $e^{\pm j\omega h/M}$ with $\omega = 3\pi/2$, $M = 8$,
as discussed in Remark \ref{Rem:intmodel}.

\begin{Remark}\label{Rem:upsampler}
For the choice of the design parameter $M$, there is a clear
trade-off between the accuracy of the internal model (and
the tracking output) and computational burden. For example,
if we increase the upsampling factor $M$ in the example above,
it is expected that the designed controller produces more accurate
sinusoids compared with the one in Fig.~\ref{Fig:controloutput},
at the expense of computational cost.
Generally speaking, our experience tells us that $M = 8$ gives a suitable
compromise.
\end{Remark}

One may also question if the above success of
hypertracking is perhaps due to the relatively ``low''
tracking signal; but it has been shown that even a
higher frequency signal of $\sin (5\pi/2)t$ can be
well tracked \cite{YYNCDC16}.

\begin{Remark} \rm
As we noted, the delay length $L$ is a design parameter.
By comparison with the case $L = 0$, we easily see that
a larger $L$ can generally provide more design freedom,
but it is not necessarily true that
a larger $L$ always leads to a better result.  This is closely
related to robustness, and will be discussed in detail
in the subsequent Section \ref{Sec:robustness}.
See also \cite{Lampeetal2003} for the behavior as we increase
$L$.
\end{Remark}

\section{Robustness}
\color{black}
\label{Sec:robustness}
In this section we discuss the robustness 
condition for hypertracking/hyperrejection problems under 
the presence of plant fluctuations or reference/disturbance frequency variations.

The following theorem clarifies the crucial relationship 
between the tracking delay $L$ and robustness:  
\begin{Thm} \label{Thm:robustness}
\em
Consider the hypertracking problem in
Fig.~\ref{Fig:1} 
with tracking/rejection signal $\sin \omega t$
and tracking delay $L = mh$.  
Under the condition of closed-loop stability, 
the closed-loop system Fig.~\ref{Fig:1} with tracking 
delay $L$ possesses 
an (approximate) internal model of this sinusoidal signal, 
(and hence robust (approximate) tracking) 
if and only if $L$ is an integer multiple 
of the period $2\pi/\omega$ of $\sin \omega t$.  
\end{Thm}
\rm

Let us first give a brief argument on this fact.  
Recall Theorem \ref{Thm:transmissionzero} on the transmission zeros
of the closed-loop system, and the remark following it.  
The tracking signal here is $\sin \omega t$, and suppose also that 
$\pm j\omega$ is not an eigenvalue of the continuous-time plant 
$P$.  Then, if tracking to $\sin \omega t$ is achieved, it means
$\lambda = e^{j\omega h}$ should be an eigenvalue of $\bar{A_{d}}$, 
and, simultaneously, $e^{j\omega h/M}$ an eigenvalue of $A_{d}$ 
as noted there (Remark \ref{Rem:intmodel}).  Hence 
with a proper choice of the hold device $H(\theta) = e^{j\omega \theta}$, 
the discrete-time controller $K(z)$ should 
work as an internal model for $\sin \omega t$.  When the 
hold device is a fast zero-order hold on $[0, h/M]$ instead, the 
tracking becomes approximate.  




{\bf Proof of Theorem \ref{Thm:robustness}}  
We adopt the framework of \cite{YYTAC94} to place 
sampled-data systems into a continuous-time scheme 
with the identification of $z \leftrightarrow e^{hs}$.  
To be more specific, the finite Laplace transform over the period 
$[0, h)$
\[
  {\bf L}[\phi](s) := \int_{0}^{h} \phi(t)e^{-st}dt 
\]
turns the discrete-time controller $K(z)$ into 
$K(e^{hs})$.  With respect to this setting, the controller 
$K(e^{hs})$ is to receive a sampled signal 
\[
  \sum_{k} e_{k}(0)e^{-khs} 
\]
which is the Laplace transform of the impulse train 
\[
    \sum_{k} e_{k}(0)\delta_{kh},
\]
where $\delta_{kh}$ is the delta distribution placed at 
point $kh$.  
The loop transfer operator then becomes 
$K(e^{hs}){\bf L}[H](s)P(s)$.  Suppose for the moment 
that $H$ is the zero-order hold.
Then ${\bf L}[H](s) = (e^{hs} - 1)/se^{hs}$, and
the loop transfer operator is expressible as
a ratio of polynomials in $s$ and $e^{hs}$.
(For a more detailed discussion, see \cite[page 710]{YYTAC94}.)
Hence this falls into the category of {\em pseudorational\/}
transfer functions  \cite{YYSIAM88,YYSIAM89}.  In fact, even when 
$H$ is not the zero-order hold, but is a compactly supported 
function on $[0, h)$, it is still pseudorational.  
They are generally expressible as ratios of entire functions of the complex 
variable $s$.  

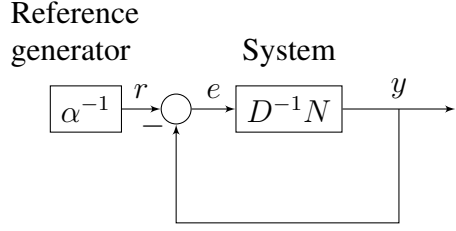
\begin{figure} [tb]
\centering
\tikzstyle{block} = [draw,  rectangle, minimum height=0.5cm, minimum width=0.4cm]
\tikzstyle{sum} = [draw, circle, node distance=1.2cm]
\tikzstyle{input} = [coordinate]
\tikzstyle{output} = [coordinate]
\tikzstyle{disturbance} = [coordinate]
\tikzstyle{pinstyle} = [pin edge={to-,thin,black}]
\begin{tikzpicture}[auto, node distance=1.5cm,>=latex']
    \node [block, name = refgen]{$\alpha^{-1}$};
    \node [sum, right of= refgen] (sum1) {};
    \node [block, right of=sum1] (system) {$D^{-1}N$};
    \node [output, right =1.5cm of system] (output){};
    \node [coordinate, below = 1.2cm of system] (belowK){};
    \node [text width=2cm, above =0.1cm of refgen]{Reference generator};
     \node [above =0.1cm of system]{System};
    \draw [->] (refgen) -- node{$r$}(sum1);
    \draw [->] (sum1) --  node{$e$}(system);
    \draw [->] (system) -- node[name=y]{$y$}(output);
    \draw [->] (y) |-(belowK) -|  node[pos=0.99]{$-$} (sum1); 
 \end{tikzpicture}
 \caption{Generalized tracking configuration}
 \label{Fig:tracking}
\end{figure}

\begin{figure}[tb]
\centering
\unitlength=0.6mm
\thicklines
\begin{picture}(130,70)(0,0)
\put(10,10){\framebox(35,20){$D^{-1}(s)N(s)$}}
\put(55,40){\framebox(40,20){$\alpha^{-1}(s)$}}

\put(75,20){\circle{10}}

\put(55,25){$y_{k}(\theta)$}
\put(80,30){$r_{k}(\theta)$}
\put(110,25){$e_{k}(\theta)$}

\put(63,15){$-$}

\put(45,20){\vector(1,0){25}}
\put(80,20){\vector(1,0){40}}
\put(75,40){\vector(0,-1){15}}

\end{picture}
\caption{Steady-state mode}
\label{Fig:steadystatetracking}
\end{figure}
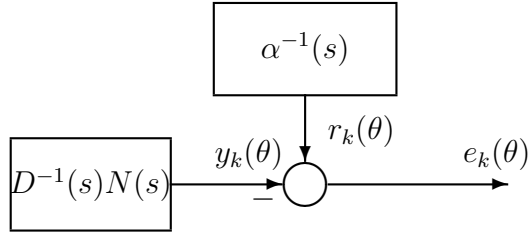

Consider the block diagram in Fig.~\ref{Fig:tracking}.  
Suppose that the tracking signal is generated by $1/\alpha(s)$ 
where $\alpha(s)$ is a polynomial in $s$.  In the present case,
$\alpha(s) = s^{2}+\omega^{2}$.  
The forward-loop system 
is described by $D^{-1}(s)N(s)$, where $D$ and $N$ are 
entire functions of exponential type.  If the tracking is
achieved, then in the steady-state mode it is equivalent to 
Fig.~\ref{Fig:steadystatetracking}.  
This is precisely in the scope of the situation considered in 
\cite[Theorem 6.4]{YYTAC94}.  Hence the asymptotic tracking 
implies that any signal generated by $\alpha^{-1}(s)$ must be 
contained in the response generated by $D^{-1}(s)$.  This
implies $\alpha(s) | D(s)$ \cite[Theorem 6.4]{YYTAC94}.  
That is, the forward loop must contain $\alpha(s)$ as an 
internal model.  If $\alpha(s) = s^{2} + \omega^{2}$ and if 
$P(s)$ does not contain $\alpha(s)$ in the denominator, then 
$\alpha(s)$ must be contained in the controller $K(z)$ 
in combination with the hold element $H(\theta)$.  

Now let us return to the issue of the tracking delay $L$.  
As we noted, the current objective is to make $y(t)$ track 
the delayed signal $r(t - L)$, not $r(t)$.  In other words, 
the argument above works ideally only for the case $L = 0$.  
When $L$ is nonzero, our design seeks a controller that 
makes $r(t - L) - y(t) \rightarrow 0$ but $e(t) = r(t) - y(t)$ 
need not go to zero.  
However, if $r(t) = \sin \omega t$ and 
if $L$ is selected to be an integer 
multiple of the period $2\pi/\omega$ of 
$\sin \omega t$, then $\sin \omega (t-L) - y(t) \rightarrow 0$ 
also implies $e(t) \rightarrow 0$.  Under this condition, we 
return to the situation discussed above, and 
$e(t) \rightarrow 0$ is guaranteed.  

Hence the above argument works again, 
and  $\alpha(s)|D(s)$, which 
means that $s^{2}+\omega^{2}$ must be included in the forward-loop 
as an internal model.  

Conversely, if the above divisibility condition 
$(2\pi/\omega)|L$ does not hold, then the tracking 
$r(t-L) - y(t) \rightarrow 0$ does not guarantee 
$e(t)\rightarrow 0$.  Indeed, $r(t-L) - y(t) \rightarrow 0$ 
and $e(t) = r(t) - y(t) \rightarrow 0$ hold simultaneously 
only when $(2\pi/\omega)|L$.  
Hence this is a necessary and sufficient condition for 
an internal model to be formed in the forward loop transfer operator.
\EndProof

\begin{Remark}\rm 
The above proof gives an argument for an ideal case 
where the internal model is given as a continuous-time 
system.  While this is possible for some choice of 
a generalized hold $H(\theta)$ (see, for example 
\cite{YYTAC94}, where $1/s$ can be generated in a 
combination of a suitable hold and a discrete-time controller), 
the internal model is not exact in the sampled-data 
context, in general.  
As we have seen in Remark \ref{Rem:intmodel}, 
the precise tracking is not achieved 
because of the finite resolution of the upsampling factor $M$, 
and the compensator cannot exactly generate the sinusoid 
$\sin \omega t$, but only approximately.  However, 
as shown in \cite{YYAutoma99}, this error due to 
sample and hold approaches the continuous-time 
internal model as $M$ increases.  
Hence the proof above shows that the result should hold in the 
limiting case.  
\end{Remark}



We give two suggestive examples.  
The nominal plant $P(s)$ is the same as (\ref{eqn:plant}): 
\[ 
 P(s) = \frac{1}{s^2+2s+1}.
\]
with the sampling time $h =1$, the upsampling factor $M=8$, 
and the delay length $L = 4h = 4$.  


\begin{Example} \rm 
 \label{Ex:robust1} 
Consider the tracking problem to the signal 
$r(t) = \sin (4\pi/3)t$.  We take the weighting function 
\[
  F_r(s) := \frac{s}{s^2+0.01s+(4\pi/3)^2 }.  
\]
\begin{figure}[tb]
\centering
    \includegraphics{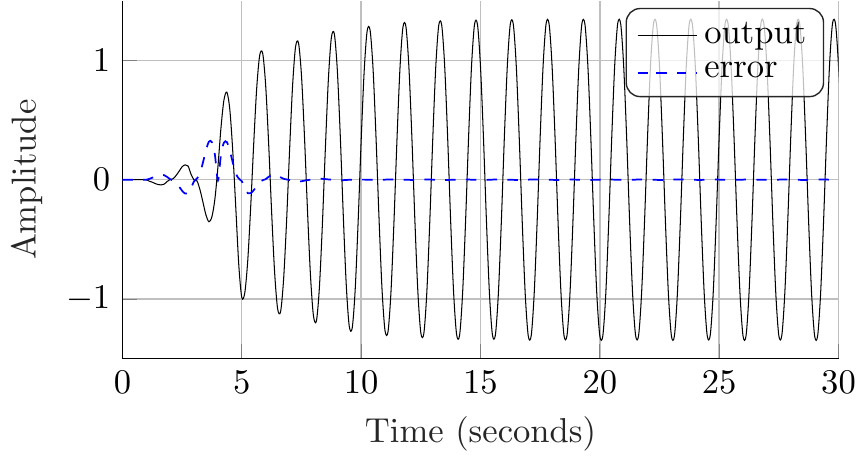}
\caption{System output tracking $\sin (4\pi/3)t$
    without disturbance} 
\label{Fig:track_4pi_ov3}
\end{figure}
As Fig.~\ref{Fig:track_4pi_ov3} shows, this gives a fine tracking 
property.  However, if we perturb the plant $P$ to 
$P + \Delta$, $\Delta(s) = 0.05/(s + 1)$, the 
resulting response exhibits a fairly large error as shown 
in Fig.~\ref{Fig:unrobustperturb}, failing to show 
a robust tracking property.  Note that the closed-loop system
remains stable here.  
\begin{figure}[t]
\centering
    \includegraphics{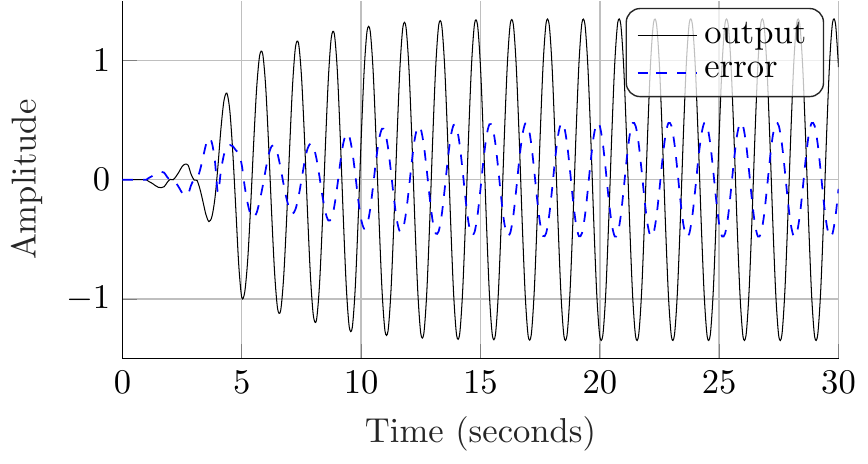}
\caption{System output for the reference $\sin (4\pi/3)t$
    under the additive perturbation $0.05/(s+1)$ to the plant} 
\label{Fig:unrobustperturb}
\end{figure}
\end{Example}

On the other hand, the following example exhibits 
quite a different behavior: 
\begin{Example} \rm \label{Ex:robust2} 
We take the same plant and simulation condition as in
Example \ref{Ex:robust1}, but with 
the tracking signal $r(t) = \sin (3\pi/2)t$.  
The result of robustness test for 
the plant perturbation 
$P \mapsto P + \Delta$, $\Delta(s) = 0.1/(s + 1)$ is 
given by Fig.~\ref{Fig:robust_trackpiov2}. 

In spite of the larger 
plant perturbation, the closed-loop system 
achieves a steady-state tracking.  
\begin{figure}[t]
\centering
    \includegraphics{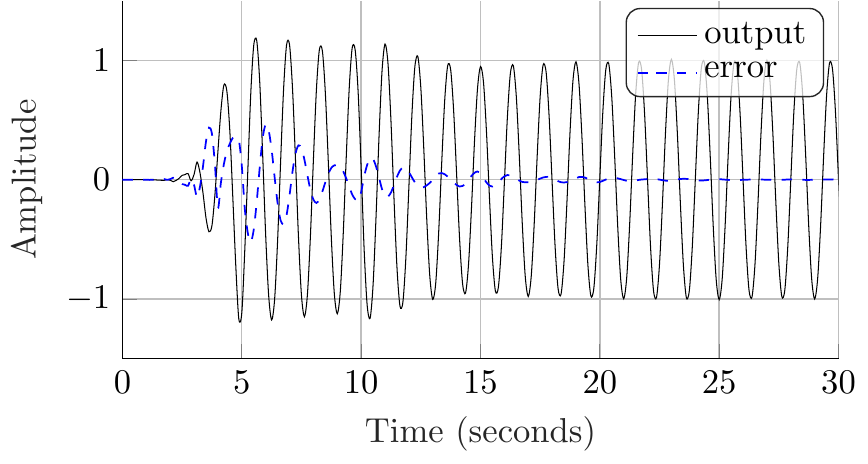}
\caption{System output tracking $\sin (3\pi/2)t$
    under the additive perturbation $0.1/(s+1)$ to the plant} 
\label{Fig:robust_trackpiov2}
\end{figure}
\end{Example}

\begin{figure}[t]
   \centering
	 \includegraphics{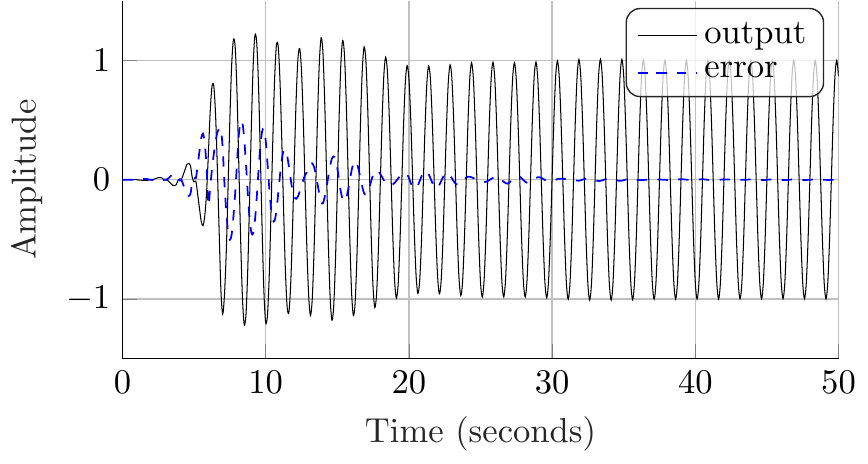}
     \caption{System output tracking $\sin (4\pi/3)t$ 
	    with $L=6$
    	under the additive perturbation $0.1/(s+1)$ to the plant} 
    \label{Fig:robust_track4piov3_delay6}
\end{figure}

In light of Theorem \ref{Thm:robustness}, the difference of 
the above two is clear:  In Example \ref{Ex:robust1}, the period 
$2\pi/(4\pi/3) = 3/2$ does not divide $L = 4$ while  
in the latter Example \ref{Ex:robust2}, $L = 4$ is an integer multiple of
$2\pi/(3\pi/2) = 4/3$, thereby assuring robust tracking.  
We can also easily ensure that taking $L = 6$ in 
Example \ref{Ex:robust1} will recover the robust tracking property
as shown in Fig.~\ref{Fig:robust_track4piov3_delay6}.  

\section{Miscellaneous Examples}
\label{Sec:examples}
\color{black}
We now give a few typical examples that can often arise in
practical situations:
\begin{itemize}
\item hypertracking for multiple sinusoids above the Nyquist frequency;
\item simultaneous tracking and disturbance rejection objectives;
\item simple hypertracking for an unstable plant;
\item hypertracking for a non-minimum phase plant.
\end{itemize}
In all the examples, the delay $L$ is chosen to be an integer multiple of
the tracking/rejection frequency so that the robustness is guaranteed.

\subsection{Hypertracking to multiple sinusoids}
\label{Subsec:multitrack}

The following example shows a case where we have two
tracking frequencies above the Nyquist frequency:

\begin{Example}\rm \label{Ex:multtrack}
(Hypertracking for multiple sinusoids)
Let
\[
  P(s) := \frac{1}{s^{2} + 2s + 1}
\]
with (normalized) sampling period $h=1$
(and hence the Nyquist frequency is $\pi$ [rad/sec].)
We aim at tracking two sinusoids
$\sin (5\pi/4)t + \sin (9\pi/4)t$, having natural frequencies
above the Nyquist frequency.  We set
the upsampling factor $M = 8$ and the delay $L =8$.
The weighting function is chosen as
\begin{equation*} \label{eqn:weight2}
  F(s) := \frac{s}{(s^2+0.01s+(5\pi/4)^2)(s^2+0.01s+(9\pi/4)^2)}
\end{equation*}
to have clear peaks at $5\pi/4$ and $9\pi/4.$
The result is shown in Fig.~\ref{Fig:TwoFreqHT3}.
We see that tracking is well achieved even for this multiple
signal tracking.
\begin{figure}[tb]
\centering
    \includegraphics{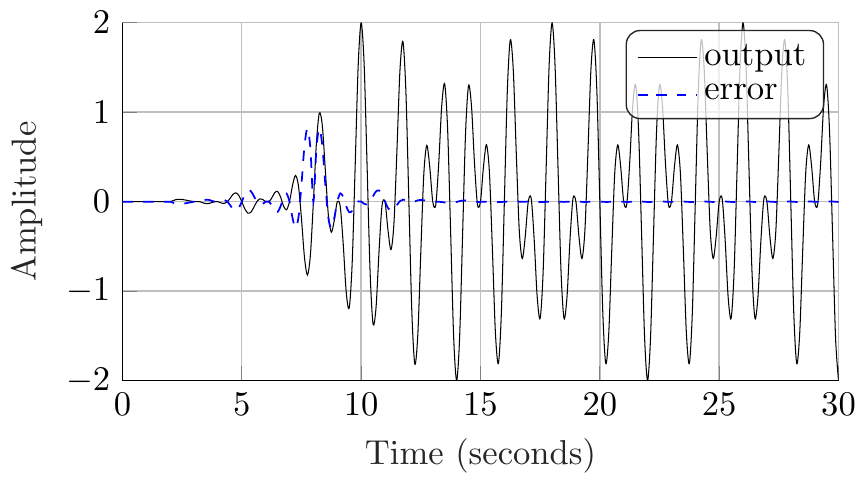} 
\caption{Hypertracking for multiple sinusoids $\sin (5\pi/4)t + \sin (9\pi/4)t$}
\label{Fig:TwoFreqHT3}
\end{figure}
\end{Example}

\subsection{Simultaneous tracking and disturbance rejection}
\label{Subsec:indist}

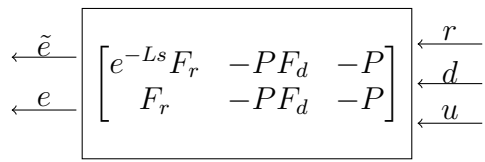
\begin{figure} [th]
\centering
\begin{tikzpicture}
\node[draw,minimum size=2cm] (GP) {$\begin{bmatrix}  e^{-Ls}F_r & -PF_d & -P\\
										F_r & -PF_d & -P \end{bmatrix}$};
\draw[arrow, <-] ([yshift=-15pt]GP.east) -- node[yshift=0.15cm]{$u$}+(1cm,0pt);
\draw[arrow, <-] (GP.east) --  node[yshift=0.15cm]{$d$}+(1cm,0pt);
\draw[arrow, <-] ([yshift=15pt]GP.east) --  node[yshift=0.15cm]{$r$} +(1cm,0pt);
\draw[arrow, <-] ([xshift=-1cm,yshift=-10pt]GP.west) -- node[yshift=0.15cm]{$e$} +(1cm,0pt);
\draw[arrow, <-] ([xshift=-1cm,yshift=10pt]GP.west) -- node[yshift=0.15cm]{$\tilde{e}$} +(1cm,0pt);
\end{tikzpicture}
 \caption{Generalized plant in the presence of input disturbance}
 \label{fig:genplantindist}
\end{figure}

We now consider the simultaneous tracking and disturbance rejection problem
given in Fig.~\ref{Fig:1}
where the disturbance $d$ is injected before the plant $P$.
The generalized plant for the design is shown in Fig.~\ref{fig:genplantindist}
where $F_r$ and $F_d$ are the weights on the reference signal and the
disturbance, respectively.

\begin{Example}\rm \label{Ex:trackreject1}
(Simultaneous tracking and rejection)
Let $h=1,$ $M = 8,$ $L =8$, and
\[
  P(s) := \frac{1}{s^{2} + 2s + 1}.
\]
Our objective here is to track $r(t)=\sin \omega_r t$
while the system is subject to the disturbance
$d(t) = \sin \omega_d t$.

We here set
$\omega_r := \pi/4$ and $\omega_d := 3\pi/2$; that is,
the tracking frequency is low, and there is a high-frequency
disturbance above the Nyquist frequency.  We commonly encounter
such a situation, e.g., in hard-disk drives, where the tracking
frequency is below the Nyquist frequency but the disturbance is
above it.
We choose the weighting functions as
\begin{equation} \label{eqn:weight3}
  F_r(s) := \frac{s}{s^2+0.01s+\omega_r^2 }, \quad
  F_d(s) := \frac{s}{s^2+0.01s+\omega_d^2}.
\end{equation}

The delayed output and the delayed error are shown
in Fig.~\ref{Fig:rejecttrack1_4pi}.
This simultaneous tracking and disturbance rejection problem is
reasonably well performed.
\begin{figure}[t]
\centering
    \includegraphics{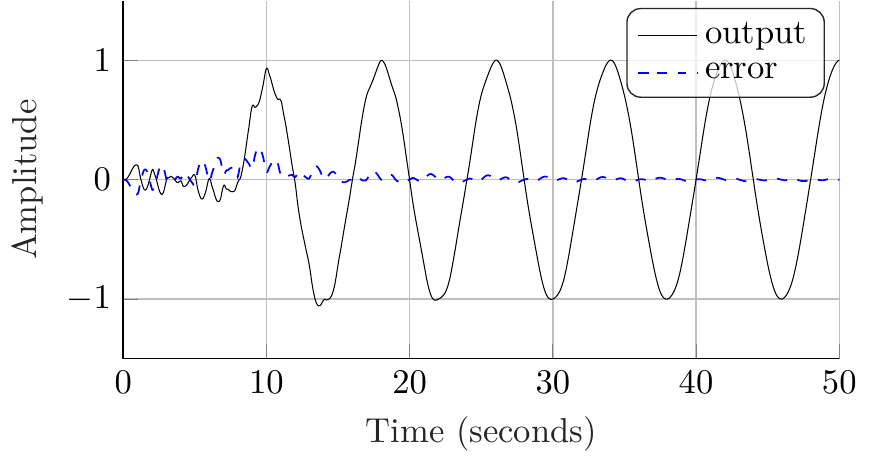}
\caption{Delayed error against $\sin (\pi/4)t$ in the presence of the input
disturbance $\sin (3\pi/2)t$}
\label{Fig:rejecttrack1_4pi}
\end{figure}
\end{Example}

The following example treats a little more delicate case where
the tracking and rejection signals are at the same frequency:

\begin{Example}\rm \label{Ex:trackreject2}
(Simultaneous tracking and rejection of the same frequency)
We now consider a more demanding case of
tracking and rejecting the same sinusoid of $\sin (3\pi/2)t$
for the same plant as in Example~\ref{Ex:trackreject1}.
The weighting functions are set to be in the form \eqref{eqn:weight3}
with $\omega_r = \omega_d = 3\pi/2.$
The delayed output and the delayed error are shown
in Fig.~\ref{Fig:HTHR_SameFreq}.
While the response is somewhat slower, the result shows good
tracking/rejection.
\begin{figure}[tb]
\centering
    \includegraphics{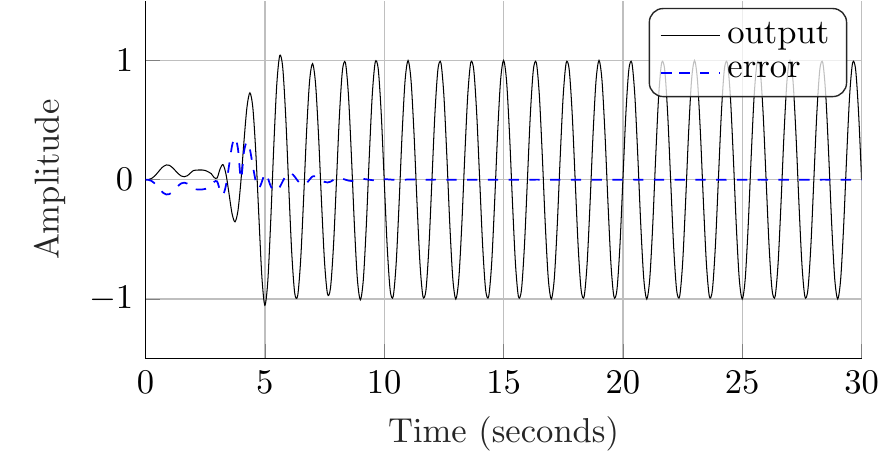}
\caption{Delayed error against $\sin (3\pi/2)t$ in the presence of the input
disturbance $\sin (3\pi/2)t$}
\label{Fig:HTHR_SameFreq}
\end{figure}
\end{Example}

\begin{Remark}
It may be noted that the transfer operator
from the disturbance to the output is {\em not\/} the
complementary sensitivity function, but rather
$P/(1 + PC)$. Therefore, the classical trade-off between
the sensitivity function and the complementary sensitivity
function in a closed-loop system does not apply here.
See also \cite{KYYYNagCCTA17} for more details in
multiple signal tracking and rejection, with a slightly different
two-step design method.
\end{Remark}

\subsection{Simple hypertracking for an unstable plant}

We have so far considered only a stable and minimum-phase plant.
We will now see that hypertracking (and hyperrejection) also works
for unstable or non-minimum phase plants.

The following example shows a case for an unstable plant:
\begin{Example}\rm \label{Ex:unstable}
(Hypertracking for an unstable plant)
Take an unstable plant $P$:
\[
  P(s) := \frac{1}{(s-0.5)},
\]
with $h = 1,$ $M=8,$ $L = 4,$ and the weighting
\begin{equation*} \label{eqn:weight5}
  F(s) = \frac{s}{s^2+0.01s+ (3\pi/2)^2}.
\end{equation*}

The delayed output and the delayed error against
$\sin(3\pi/2 t)$ are shown
in Fig.~\ref{Fig:HT_Unstable}.
Again, hypertracking is well achieved for this case also.
\begin{figure}[tb]
\centering
    \includegraphics{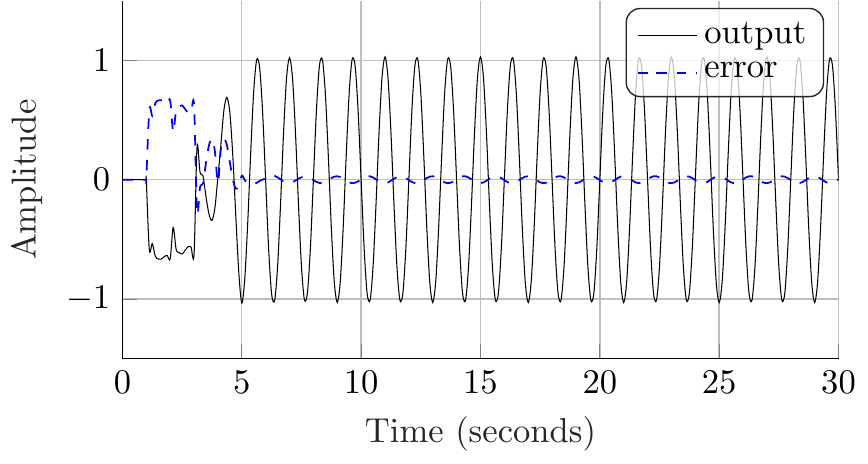}
\caption{Delayed tracking for $\sin (3\pi/2)t$ for an unstable plant}
\label{Fig:HT_Unstable}
\end{figure}
\end{Example}

\subsection{Simple hypertracking for a non-minimum phase plant}

Finally, we give the following example dealing with a non-minimum
phase plant.
\begin{Example}\rm \label{Ex:nonminimumphase}
(Hypertracking for a non-minimum phase plant.)
Take the following plant that has an unstable zero
at $s = 1$:
\[
  P(s) := \frac{s-1}{s^{2} + 2s + 1}.
\]
The tracking frequency is $3\pi/2$ as before, and
we take the weighting
\begin{equation*} \label{eqn:weight6}
  F(s) = \frac{s}{s^2+0.01s+ (3\pi/2)^2},
\end{equation*}
with the same $h = 1,$ $M = 8,$ and $L = 4$ as above.

The delayed output and the delayed error are shown
in Fig.~\ref{Fig:NMPP_HT}.
While there remain some errors, the overall tracking must
be satisfactory.
\begin{figure}[tb]
\centering
    \includegraphics{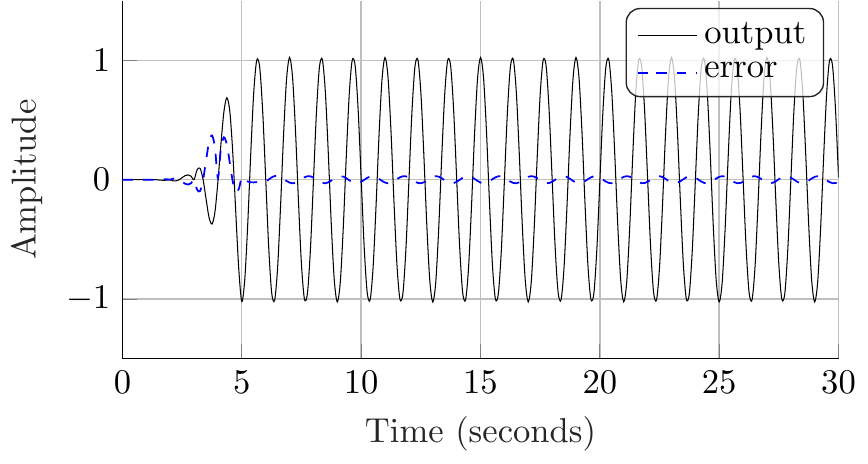}
\caption{Delayed error against $\sin (3\pi/2)t$ for a non-minimum phase plant}
\label{Fig:NMPP_HT}
\end{figure}
\end{Example}

\section{Concluding remarks}
  \label{Sec:conclusion}
We have proposed a new scheme (along with \cite{YYNCDC16}) 
for tracking/rejection of signals 
that reside either beyond or below the Nyquist frequency.
This has been made possible 
by introducing a suitable choice of signal weighting.  
When the tracking/rejection signal is above the Nyquist 
frequency, an appropriate choice allows us to control 
high-frequency intersample response.  

We have also completely characterized robustness in 
this context.  That is, the designed closed-loop 
system achieves robust tracking/rejection 
if and only if the tracking/rejection delay $L$ 
is an integer multiple of the periods of target signals 
(Theorem \ref{Thm:robustness}).  This leads to an 
interesting observation.  

In general, when there is no feedback loop, i.e., in the
case of a delayed signal reconstruction, a longer delay 
length is always advantageous; for example, in the
signal reconstruction, the designed filter will approach 
an ideal filter as $L \rightarrow \infty$ \cite{ChenFrancissignal95}. 
In the present setting, however, a longer delay does not 
necessarily yield a desirable result in view of robustness.  
A delay incompatible with the target signal period can 
behave very poorly when there is a small amount of perturbations.  

Multirate sampled-data control has been studied 
in the control literature: see, e.g., 
\cite{ArakiYamamotoTAC86,HagiwaraArakiTAC88,MitaChida88,MitaChidaTAC90}.  
However, the emphasis there is mainly on how one 
can obtain full information by multirate sampling of
the output, thereby extending the capability of control.  
It is to be noted that we do {\em not\/}  perform further sampling 
on the sampled output, and the basic sampling period 
remains intact for outputs.  Upsampling 
is performed only on the side of the control signals, 
and we focus our attention on how it can enhance 
control capability.  This is made possible by a proper
choice of weighting on tracking/rejection signals.

\bibliographystyle{IEEEtranS}

\bibliography{newyyref}

\end{document}